\documentclass[structabstract]{aa} 
\usepackage[utf8]{inputenc}
\usepackage{txfonts} 
\usepackage{graphicx} 
\usepackage{xcolor}
\usepackage{natbib} 
\usepackage{url} 

\pdfpageattr {/Group << /S /Transparency /I true /CS /DeviceRGB>>}

\bibpunct{(}{)}{;}{a}{}{,} 

\newcommand{\ie}{{\em i.e.}} 
\newcommand{\eg}{{\em e.g.}}

\newcommand{\emm}[1]{\ensuremath{#1}}
\newcommand{\emr}[1]{\emm{\mathrm{#1}}}
\newcommand{\chem}[1]{\emr{\,#1}} 
\newcommand{\unit}[1]{\emr{\,#1}}

\newcommand{\pc}{\unit{pc}} 
\newcommand{\mum}{\unit{\mu m}}

\newcommand{\pscm}{\unit{cm^{-2}}} 

\newcommand{\kms}{\unit{km\,s^{-1}}}

\newcommand{\K}{\unit{K}}
\newcommand{\GHz}{\unit{GHz}} 
\newcommand{\Kkms}{\unit{K\,km\,s^{-1}}}
\newcommand{\asinh}{\unit{asinh}} 
\newcommand{\thCO}{\chem{^{13}CO}}
\newcommand{\twCO}{\chem{^{12}CO}} 
\newcommand{\CeiO}{\chem{C^{18}O}}
\newcommand{\Jone}{(1--0)}

\newcommand{\Ht}{\emr{H_2}} 
\newcommand{\nH}{\emm{n_\emr{H}}}
 
\newcommand{\NHt}{\emm{N(\Ht)}}
\newcommand{\U}{\emm{U}}

\newcommand{\Hii}{\ion{H}{ii}} 
\newcommand{\dix}[1]{\emm{10^{#1}}}

\definecolor{ochre}{rgb}{0.8, 0.47, 0.13}

\newcommand{\FigData}{ 
\begin{figure*}
	\centering
	\includegraphics[width=0.98\linewidth]{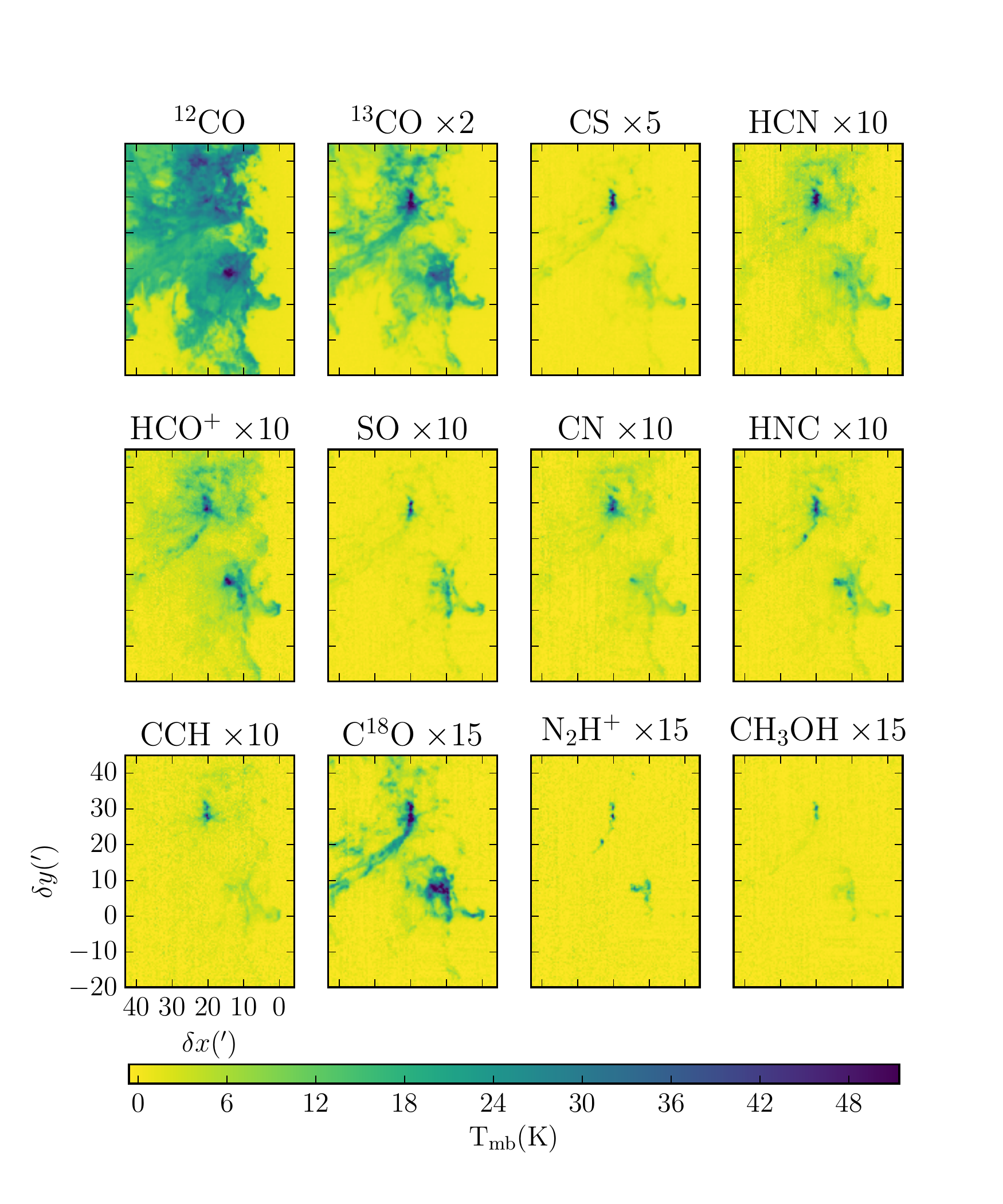}
	\caption{Maps of molecular emission in Kelvin main beam temperatures.}
	\label{fig:data} 
\end{figure*}
}

\newcommand{\FigDataAsinh}{ 
\begin{figure*}
	\centering
	\includegraphics[width=0.98\linewidth]{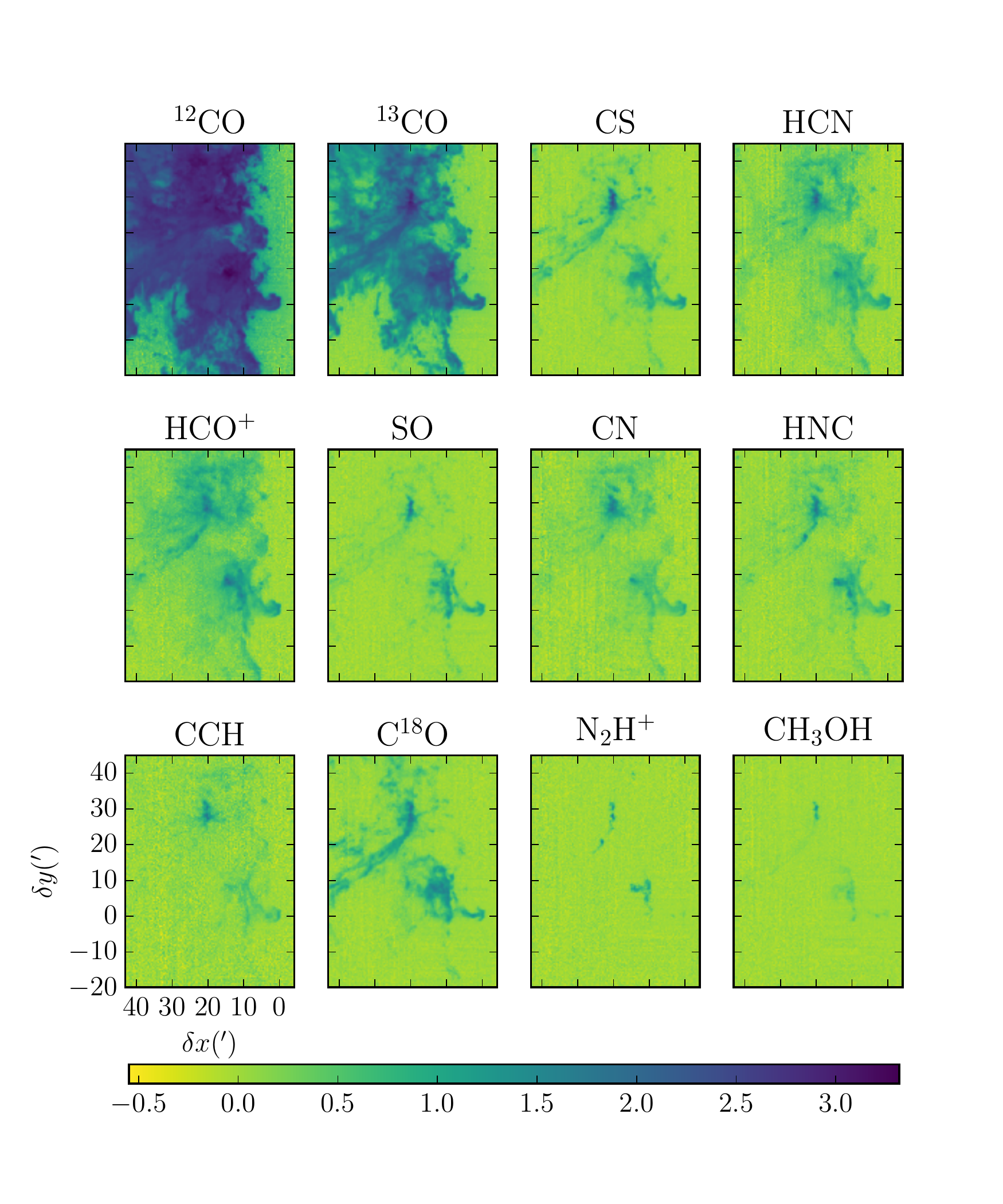} 
	\caption{Maps
	of molecular emission after reparametrization by the asinh function.}
	\label{fig:data_asinh} 
\end{figure*}
}

\newcommand{\FigAsinh}{ 
\begin{figure}
	\centering
	\includegraphics[width=\linewidth]{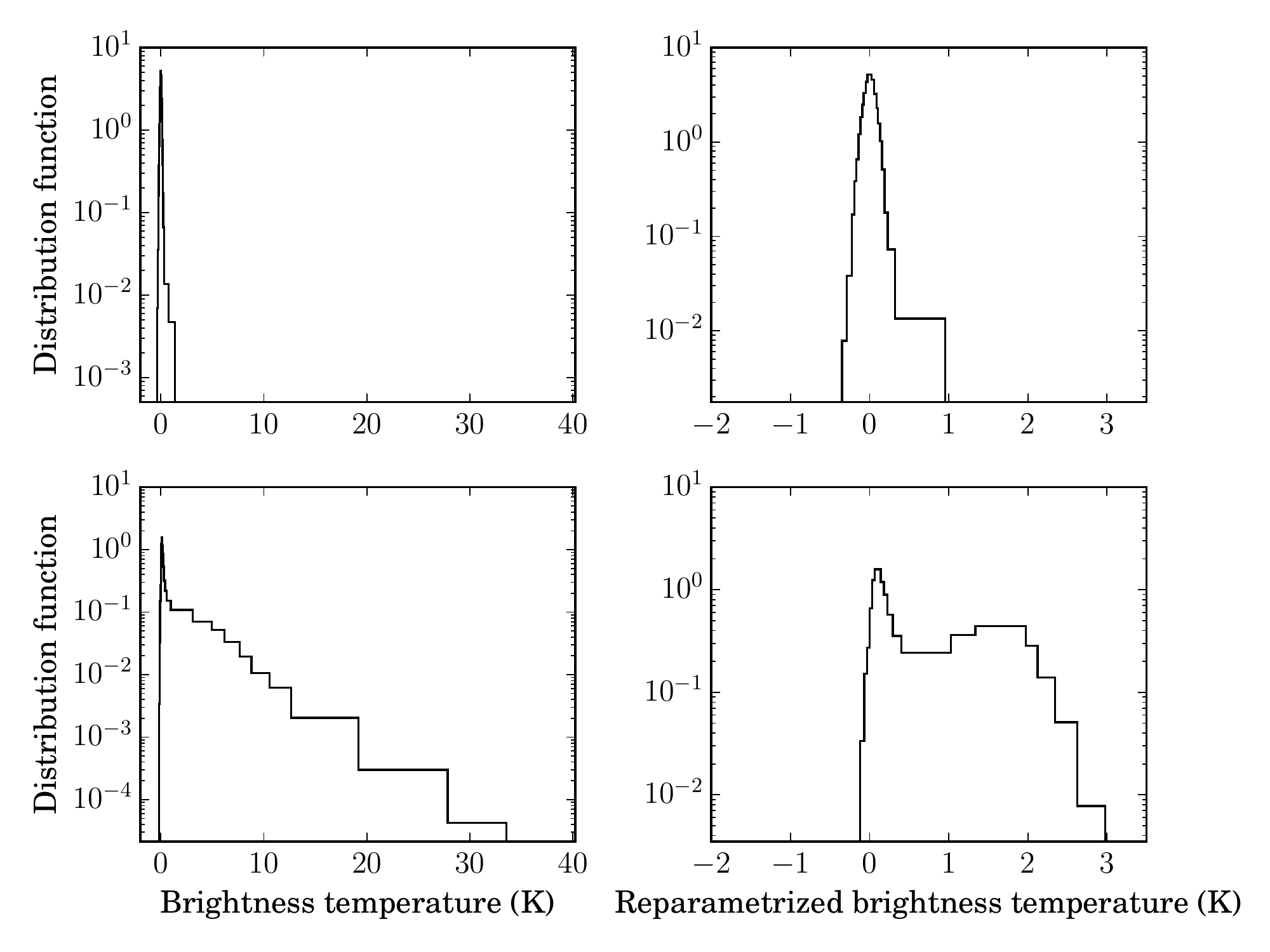} 
	\caption{Effect of the
	asinh renormalization on the intensity distributions. \emph{Left
	column:} before renormalization, \emph{right column:} after
	renormalization, \emph{Top row:} \chem{N_2H^+}, \emph{bottom row:}
	\thCO.} 
	\label{fig:asinh_reparam} 
\end{figure}
}

\newcommand{\FigAsinhFunction}{ 
\begin{figure}
	\centering
	\includegraphics[width=\linewidth]{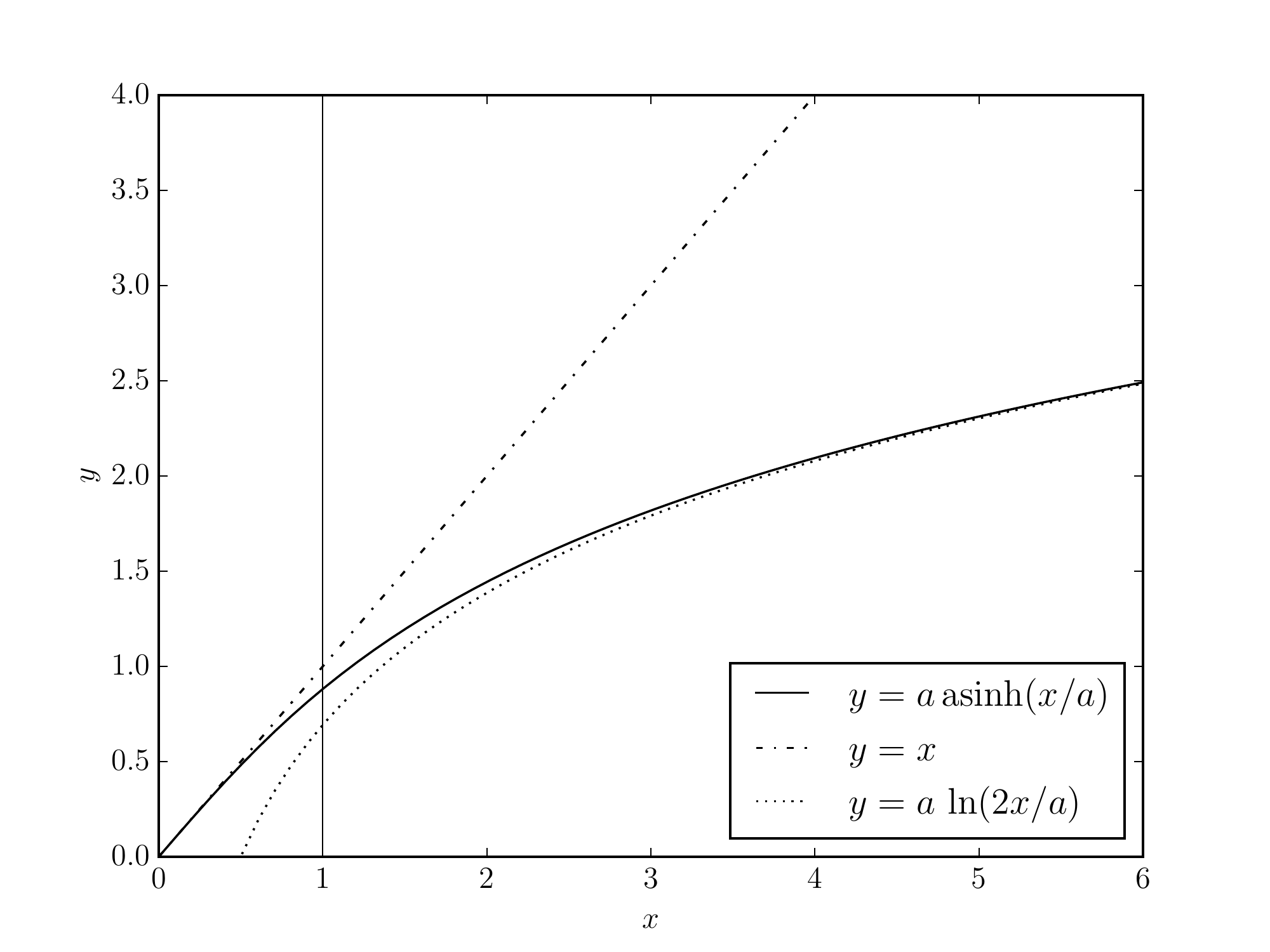} 
	\caption{Plot
	of the $\asinh$ function (solid line) showing the asymptotes when
	$x\rightarrow0$ (dash dotted line) and $x\rightarrow+\infty$ (dotted
	line). The parameter $a$ (here $a=1$) is traced with a thin vertical
	line} 
	\label{fig:asinh_function} 
\end{figure}
}

\newcommand{\FigProjMaps}{ 
\begin{figure*}
	\centering
	\includegraphics[width=0.96\linewidth]{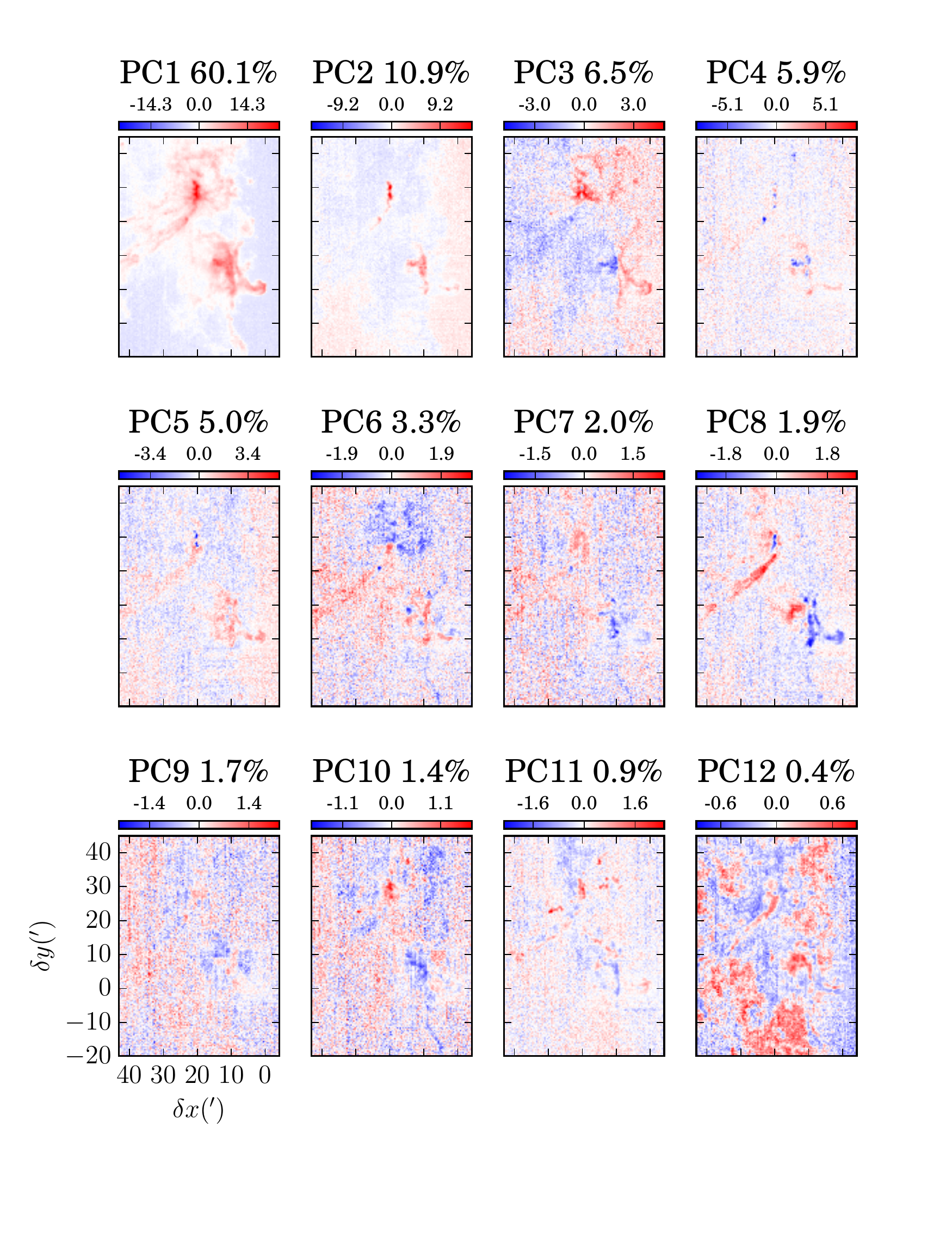}
	\caption{Principal component maps. These maps represent the value of
	each observed pixel when they are projected in the space of the
	principal components.} 
	\label{fig:proj_maps} 
\end{figure*}
}

\newcommand{\FigExplainedVariance}{ 
\begin{figure}
	\centering
	\includegraphics[width=\linewidth]{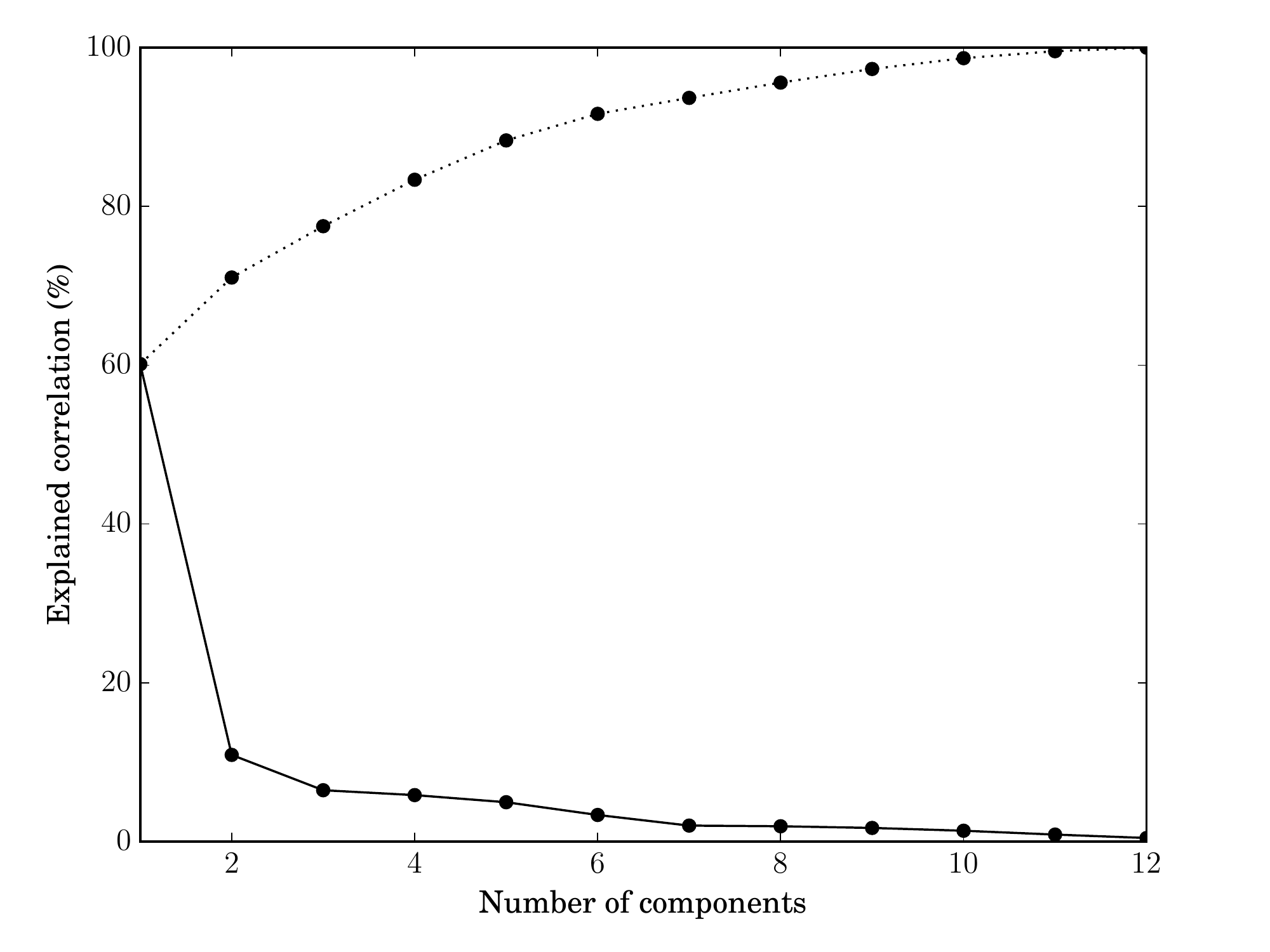}
	\caption{Percentage of the explained correlation as a function of the
	number of components in the Principal Component Analysis, \emph{dotted
	line:} cumulative percentage.
	} 
	\label{fig:explained_variance} 
\end{figure}
}

\newcommand{\FigWheel}{ 
\begin{figure*}
	\centering
	\includegraphics[width=\linewidth]{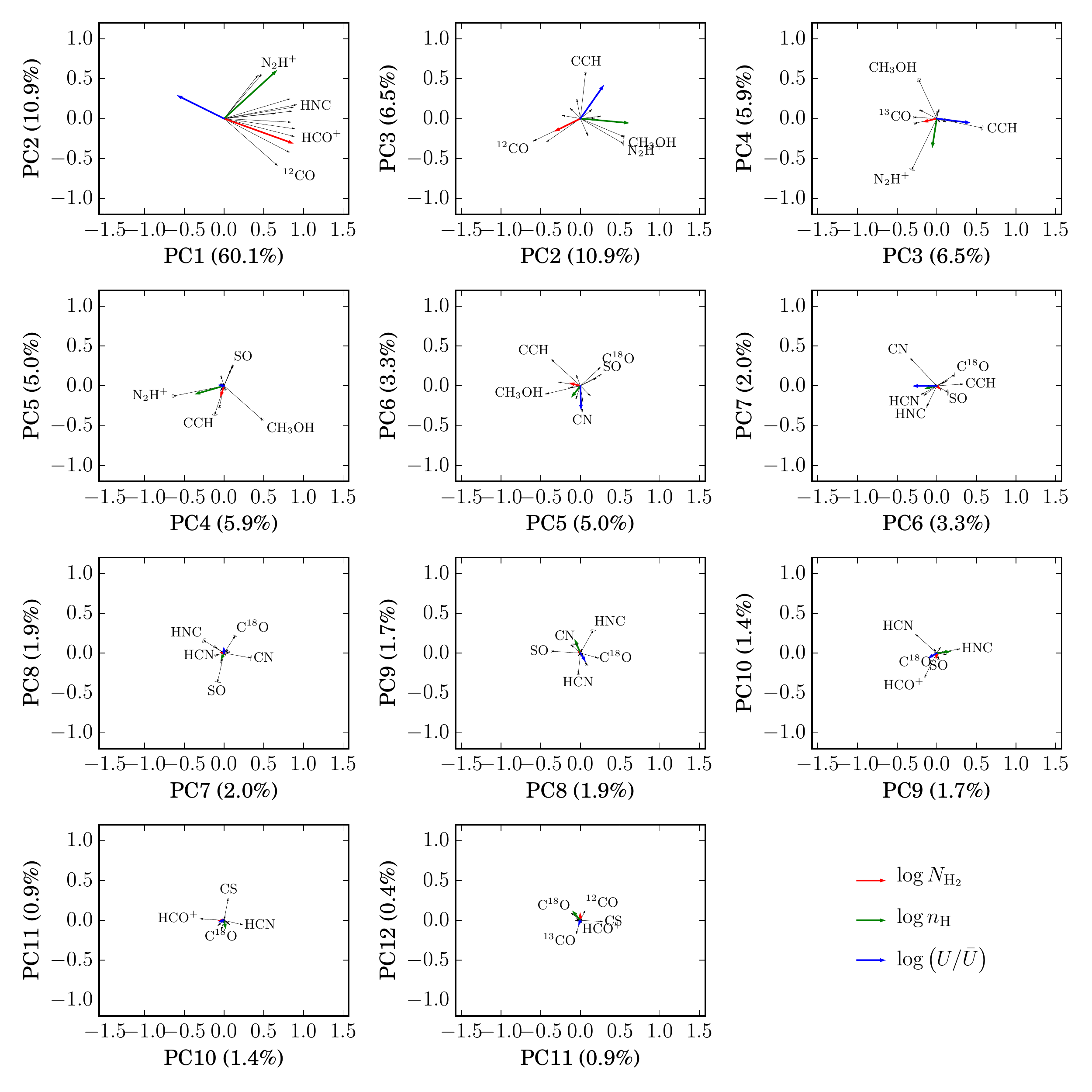}
	\caption{{Correlation wheels, showing the initial line
	intensities as vector having as coordinates their correlation
	coefficients to each PC, represented in the planes of successive pairs
	of PCs. Uncertainties from our bootstraping analysis (see
	Sect.~\ref{sec:noise}) are presented as thin black contours around the
	arrows' heads (isodensity contours containing 68\% of the distribution).
	Also represented in colored arrows are the correlations of our
	independent physical parameters with the PCs (red: $\log
	N_{\mathrm{H_2}}$, green: $\log n_{\mathrm{H}}$, blue: $\log
	\left(U/\bar{U}\right)$).}} 
	\label{fig:wheel} 
\end{figure*}
}

\newcommand{\FigEigs}{ 
\begin{figure*}
	\centering
	\includegraphics[width=\linewidth]{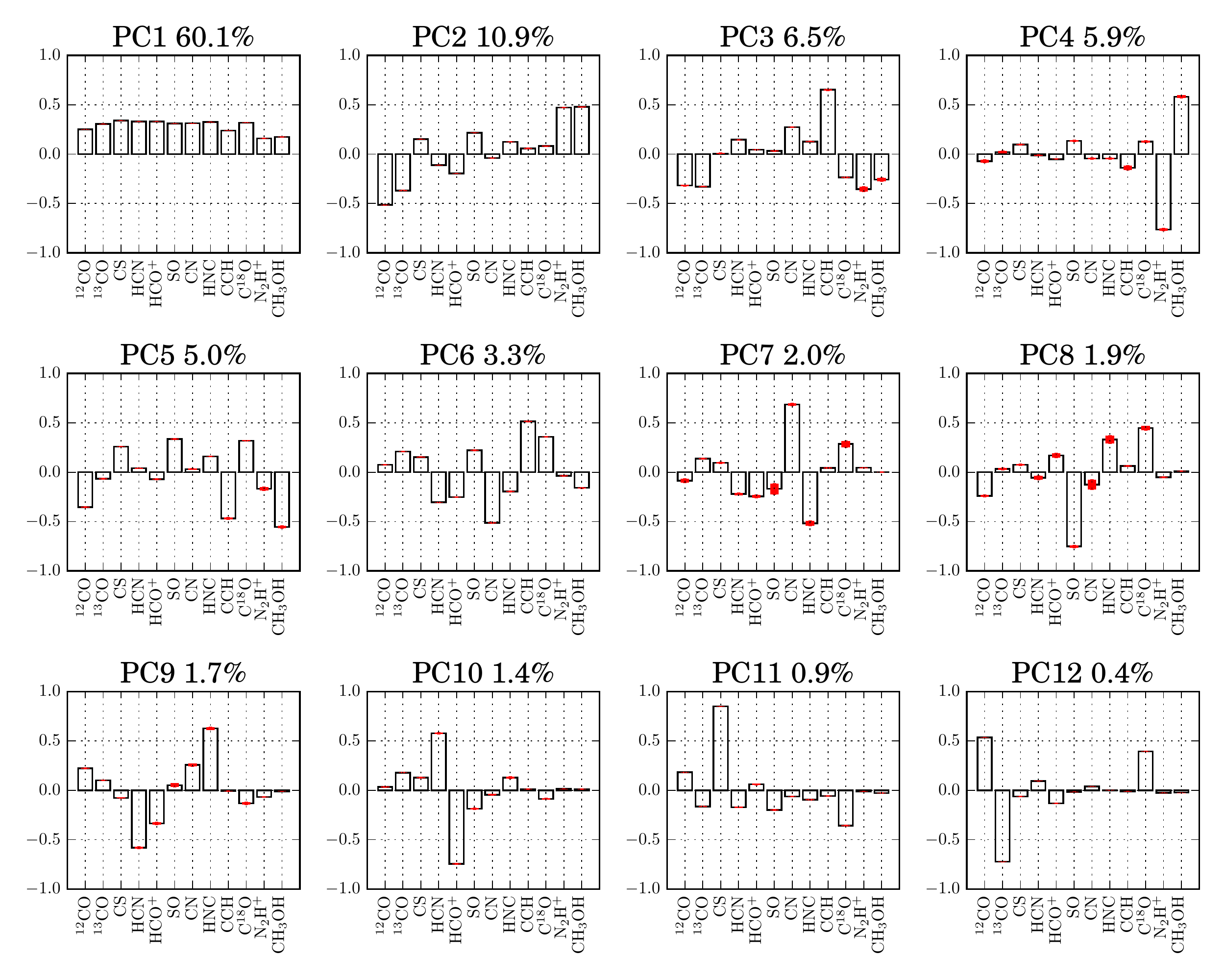}
	\caption{{Bar plots showing the contribution of each line
	intensity to each principal component (with the fraction of the total
	correlation accounted for by each PC given as a percentage). The
	uncertainties (standard deviations) shown in red are obtained by
	bootstrapping as described in Sect.~\ref{sec:noise}}.} 
	\label{fig:eigs}
\end{figure*}
}

\newcommand{\FigPhysicalParams}{ 
\begin{figure*}
	\centering
	\includegraphics[width=\linewidth,trim={0 3cm 0
	3cm}]{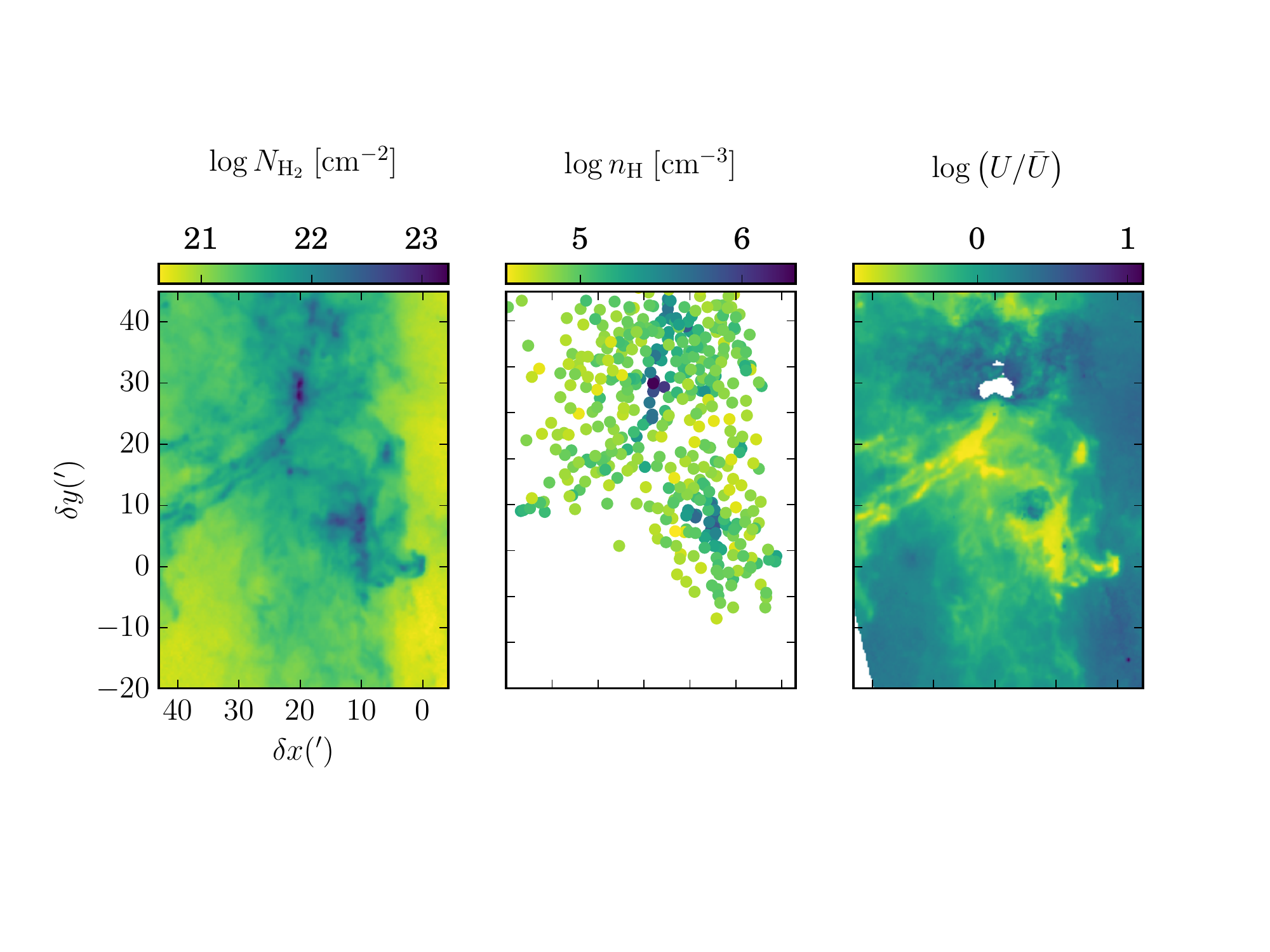} 
	\caption{Maps of the independently
	measured physical parameters, $H_2$ column density (\emph{left}),
	volumic density (\emph{middle}) , UV illumination (\emph{right}) }
	\label{fig:physical_params} 
\end{figure*}
}

\newcommand{\FigPhysicalCorr}{ 
\begin{figure*}
	\centering
	\includegraphics[width=\linewidth]{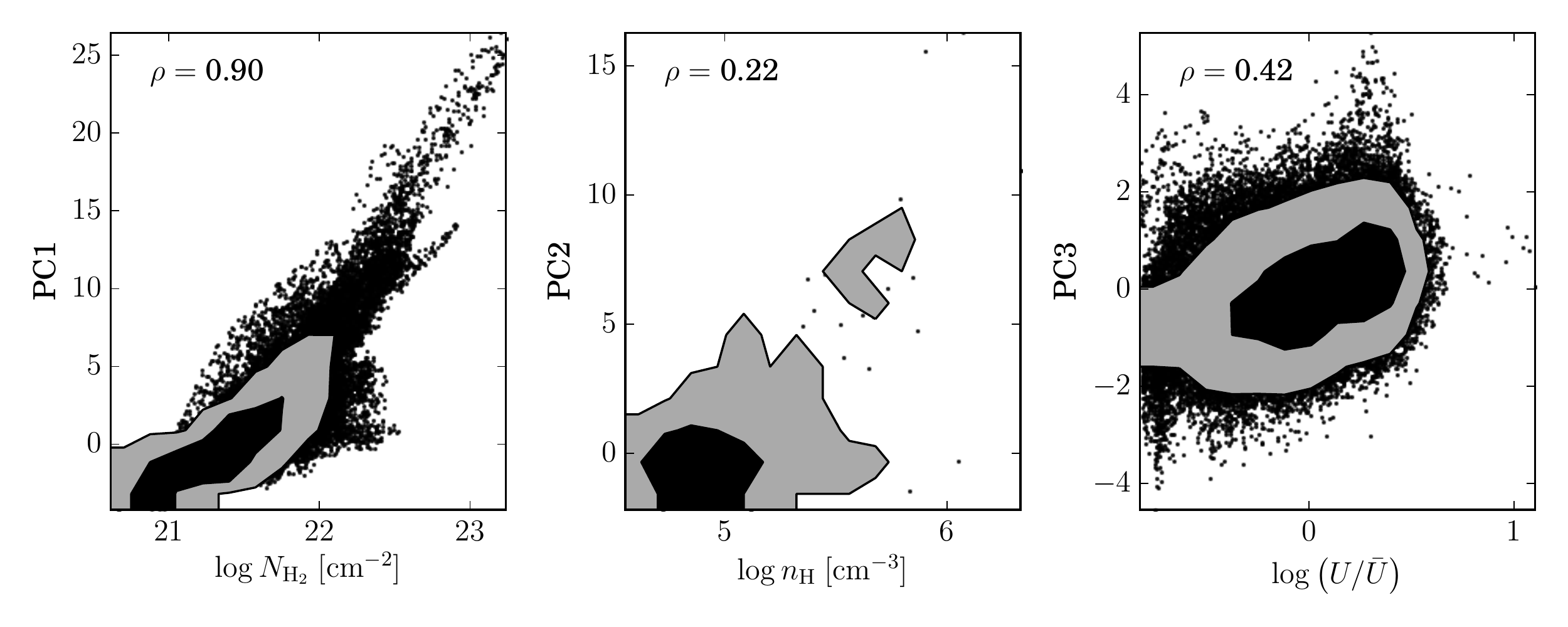}
	\caption{Scatter plots of the first three principal components with the
	independent physical parameters. Contours in black and gray correspond
	to 68\% and 95\% of the samples respectively.} 
	\label{fig:correl}
\end{figure*}
}

\newcommand{\FigColor}{ 
\begin{figure*}
	\centering
	\includegraphics[width=0.98 \linewidth]{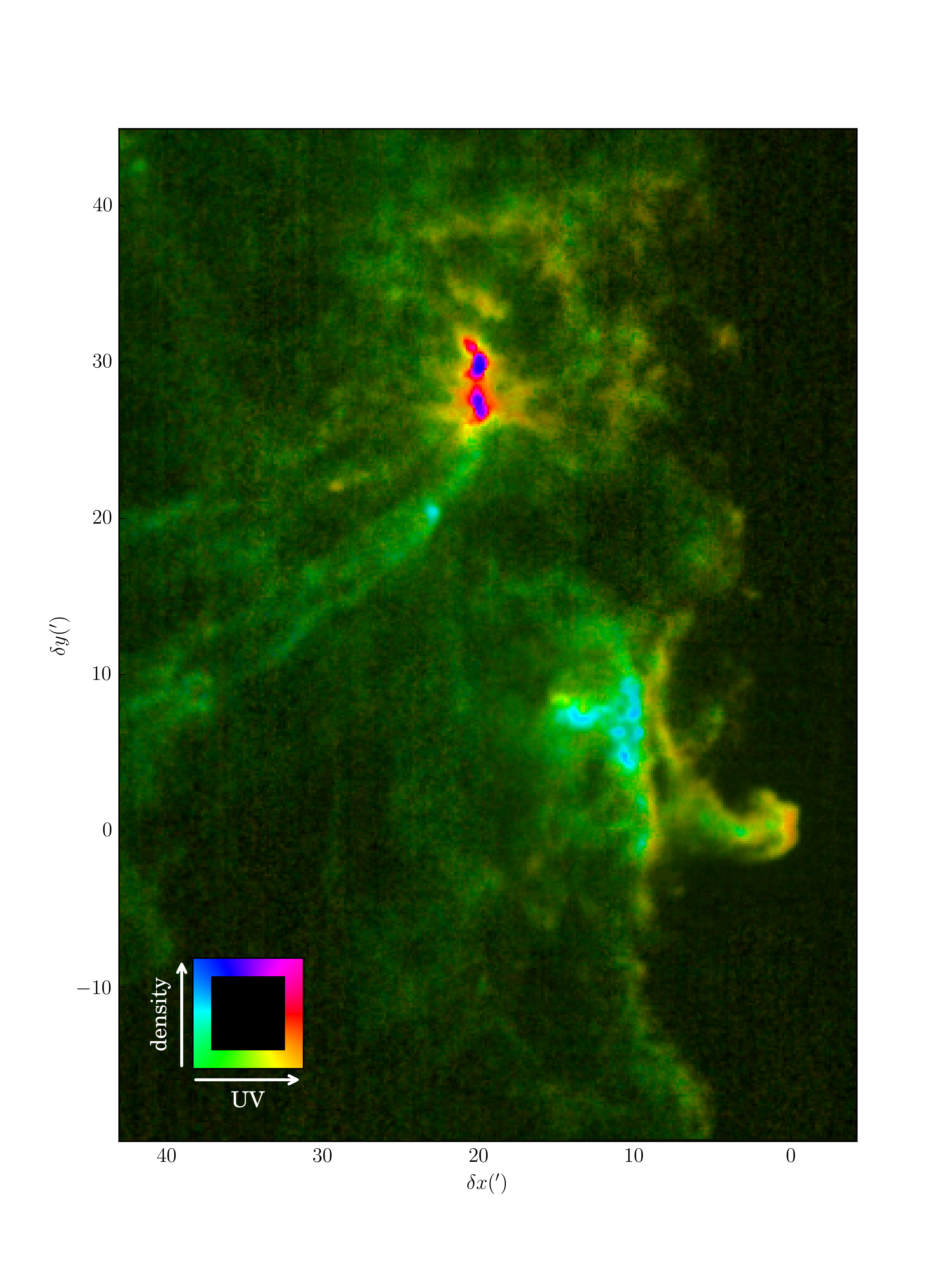}
	\caption{Synthetic view of the Orion\,B molecular cloud. In this
	colormap, the intensity of each pixel is encoded by PC1 (column density)
	and the hue is encoded by the angle of the vector constructed using two
	orthogonal components PC2 (density) and PC3 (UV radiation field). It is
	possible to identify limiting cases. Magenta:
	dense PDR, yellow: diffuse PDR, green: diffuse non illuminated, blue:
	dense non illuminated} 
	\label{fig:color} 
\end{figure*}
}

\newcommand{\FigOptimalK}{ 
\begin{figure}
	\centering
	\includegraphics[width=\linewidth]{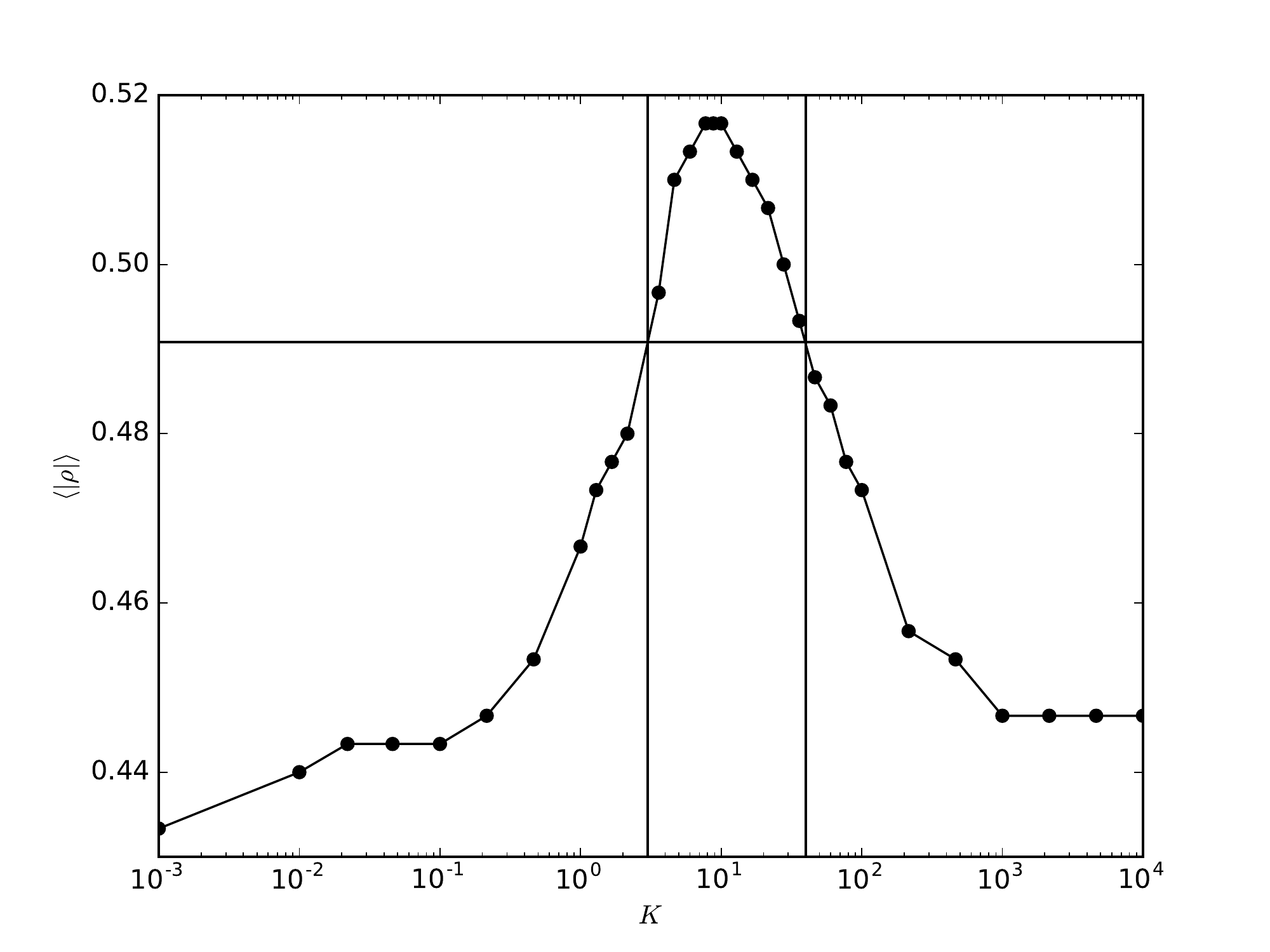} 
	\caption{Variation
	of $\langle|\rho|\rangle$, the mean of the absolute values of the
	Spearman's correlation coefficients as a function of $K$, with $a=K
	\emr{median}(\sigma)$. The optimal value is found for $K=8$.}
	\label{fig:optimalK} 
\end{figure}
}

\begin{document}

\title{Dissecting the molecular structure of the Orion\,B cloud: \\
Insight from Principal Component Analysis\thanks{Based on observations
carried out at the IRAM-30m single-dish telescope. IRAM is supported by
INSU/CNRS (France), MPG (Germany) and IGN (Spain).}} 

\author{Pierre Gratier\inst{\ref{LAB}} 
\and Emeric
Bron\inst{\ref{ICMM},\ref{LERMA_Meudon}} 
\and Maryvonne
Gerin\inst{\ref{LERMA_ENS}} 
\and Jérôme
Pety\inst{\ref{IRAM},\ref{LERMA_ENS}} 
\and Viviana V.
Guzman\inst{\ref{CFA}} 
\and Jan Orkisz\inst{\ref{IRAM},\ref{LERMA_ENS}}
\and S\'ebastien Bardeau\inst{\ref{IRAM}} 
\and Javier R.
Goicoechea\inst{\ref{ICMM}} 
\and Franck Le
Petit\inst{\ref{LERMA_Meudon}} 
\and Harvey Liszt\inst{\ref{NRAO}} 
\and
Karin \"Oberg\inst{\ref{CFA}} 
\and Nicolas Peretto\inst{\ref{UC}} 
\and
Evelyne Roueff\inst{\ref{LERMA_Meudon}} 
\and Albrech
Sievers\inst{\ref{IRAM}} 
\and Pascal Tremblin\inst{\ref{CEA}}}

\institute{ Laboratoire d'astrophysique de Bordeaux, Univ. Bordeaux,
CNRS, B18N, allée Geoffroy Saint-Hilaire, 33615 Pessac, France
\label{LAB} 
\and ICMM, Consejo Superior de Investigaciones Cientificas
(CSIC). E-28049. Madrid, Spain.
\label{ICMM} 
\and LERMA, Observatoire de Paris, PSL Research University,
CNRS, Sorbonne Universités, UPMC Univ. Paris 06, F-92190, Meudon,
France 
\label{LERMA_Meudon} 
\and LERMA, Observatoire de Paris, PSL
Research University, CNRS, Sorbonne Universités, UPMC Univ. Paris 06,
École normale supérieure, F-75005, Paris, France 
\label{LERMA_ENS}
\and IRAM, 300 rue de la Piscine, 38406 Saint Martin d'Hères, France
\label{IRAM} 
\and Harvard-Smithsonian Center for Astrophysics, 60 Garden
Street, Cambridge, MA, 02138, USA.
\label{CFA} 
\and National Radio Astronomy Observatory, 520 Edgemont
Road, Charlottesville, VA, 22903, USA 
\label{NRAO} 
\and School of
Physics and Astronomy, Cardiff University, Queen's buildings, Cardiff
CF24 3AA, UK.
\label{UC} 
\and Maison de la Simulation, CEA-CNRS-INRIA-UPS-UVSQ, USR
3441, Centre d'étude de Saclay, F-91191 Gif-Sur-Yvette, France.
\label{CEA} } 

\date{} 

\abstract {The combination of wideband receivers and spectrometers
currently available in (sub-)millimeter observatories deliver wide-field
hyperspectral imaging of the interstellar medium. Tens of spectral lines
can be observed over degree wide fields in about fifty hours. This
wealth of data calls for restating the physical questions about the
interstellar medium in statistical terms.} {We aim at gaining
information on the physical structure of the interstellar medium from a
statistical analysis of many lines from different species over a large
field of view, without requiring detailed radiative transfer or
astrochemical modeling.} {We coupled a nonlinear rescaling of the data
with one of the simplest multivariate analysis methods, namely the
Principal Component Analysis, to decompose the observed signal into
components that we interpret first qualitatively and then quantitatively
based on our deep knowledge of the observed region and of the
astrochemistry at play.} {We identify 3 principal components, linear
compositions of line brightness temperatures, that are correlated at
various levels with the column density, the volume density and the UV
radiation field.} {When sampling a sufficiently diverse mixture of
physical parameters, it is possible to decompose the molecular emission
in order to gain physical insight on the observed interstellar medium.
This opens a new avenue for future studies of the interstellar medium.}


\maketitle{} 

\section{Introduction}

\begin{table*}
	\caption{ 
	\label{tab:value}
	Properties of the observed spectral lines. The last six columns show the
	statistics of the data before and after asinh reparametrization.}
	\begin{center}
		{\scriptsize 
		\begin{tabular}
			{llrr|rrrr|rrrr} \hline & &
			& & \multicolumn{4}{c}{Original data} & \multicolumn{4}{c}{After asinh
			reparametrization}\\
			Molecule                                 & Transitions                                                                                          & Frequency                                & Noise    & Min.     & Median   & Max.     & Std.     & Min.     & Median   & Max.     & Std.     \\ 
                                         &                                                                                                      & (MHz)                                    & (K)      &  (K)       &   (K)     &   (K)      &    (K)     &  (K)       &   (K)      &   (K)      &    (K)     \\ 
\hline
$\ensuremath{\mathrm{^{12}CO}}$          & $\ensuremath{\mathrm{J=1 \rightarrow 0}}$                                                            &  115271.202 & $  0.09$ & $ -0.39$ & $ 13.40$ & $ 57.11$ & $ 10.18$  & $ -0.37$ & $  2.39$ & $  3.32$ & $  0.91$ \\ 
$\ensuremath{\mathrm{^{13}CO}}$          & $\ensuremath{\mathrm{J=1 \rightarrow 0}}$                                                            &  110201.354 & $  0.04$ & $ -0.19$ & $  1.38$ & $ 36.43$ & $  3.27$  & $ -0.19$ & $  0.97$ & $  3.03$ & $  0.74$ \\ 
$\ensuremath{\mathrm{CS}}$               & $\ensuremath{\mathrm{J=2 \rightarrow 1}}$                                                            &  97980.953 & $  0.06$ & $ -0.36$ & $  0.06$ & $ 15.53$ & $  0.48$  & $ -0.35$ & $  0.06$ & $  2.48$ & $  0.23$ \\ 
$\ensuremath{\mathrm{HCN}}$              & $\ensuremath{\mathrm{J=1 \rightarrow 0,\, F=2 \rightarrow 1}}$                                       &  88631.848 & $  0.10$ & $ -0.58$ & $  0.15$ & $ 10.32$ & $  0.39$  & $ -0.52$ & $  0.15$ & $  2.22$ & $  0.25$ \\ 
$\ensuremath{\mathrm{HCO^+}}$            & $\ensuremath{\mathrm{J=1 \rightarrow 0}}$                                                            &  89188.525 & $  0.09$ & $ -0.45$ & $  0.26$ & $  8.07$ & $  0.47$  & $ -0.42$ & $  0.25$ & $  2.07$ & $  0.30$ \\ 
$\ensuremath{\mathrm{SO}}$               & $\ensuremath{\mathrm{N=3 \rightarrow 2,\, J=2 \rightarrow 1}}$                                       &  99299.870 & $  0.06$ & $ -0.43$ & $  0.04$ & $  6.46$ & $  0.24$  & $ -0.40$ & $  0.04$ & $  1.92$ & $  0.17$ \\ 
$\ensuremath{\mathrm{CN}}$               & $\ensuremath{\mathrm{N=1 \rightarrow 0,\, J=3/2 \rightarrow 1/2,\, F=5/2 \rightarrow 3/2 }}$         &  113490.970 & $  0.09$ & $ -0.58$ & $  0.10$ & $  6.33$ & $  0.27$  & $ -0.52$ & $  0.09$ & $  1.91$ & $  0.20$ \\ 
$\ensuremath{\mathrm{HNC}}$              & $\ensuremath{\mathrm{J=1 \rightarrow 0}}$                                                            &  90663.568 & $  0.08$ & $ -0.49$ & $  0.07$ & $  6.01$ & $  0.27$  & $ -0.45$ & $  0.07$ & $  1.88$ & $  0.19$ \\ 
$\ensuremath{\mathrm{CCH}}$              & $\ensuremath{\mathrm{N=1 \rightarrow 0,\, J=3/2 \rightarrow 1/2,\, F=2 \rightarrow 1}}$              &  87316.898 & $  0.12$ & $ -0.62$ & $  0.08$ & $  5.72$ & $  0.22$  & $ -0.55$ & $  0.08$ & $  1.85$ & $  0.18$ \\ 
$\ensuremath{\mathrm{C^{18}O}}$          & $\ensuremath{\mathrm{J=1 \rightarrow 0}}$                                                            &  109782.173 & $  0.06$ & $ -0.30$ & $  0.06$ & $  5.55$ & $  0.42$  & $ -0.29$ & $  0.06$ & $  1.83$ & $  0.26$ \\ 
$\ensuremath{\mathrm{N_2H^+}}$           & $\ensuremath{\mathrm{J=1 \rightarrow 0,\, F1=2 \rightarrow 1,\, F=3 \rightarrow 2 }}$                &  93173.764 & $  0.08$ & $ -0.44$ & $  0.00$ & $  4.53$ & $  0.13$  & $ -0.41$ & $  0.00$ & $  1.70$ & $  0.10$ \\ 
$\ensuremath{\mathrm{CH_3OH}}$           & $\ensuremath{\mathrm{J=2 \rightarrow 1,\, K=0 \rightarrow 0,\, (A+)}}$                               &  96741.375 & $  0.06$ & $ -0.34$ & $  0.01$ & $  2.24$ & $  0.08$  & $ -0.32$ & $  0.01$ & $  1.26$ & $  0.08$ \\ 
 \hline 
		\end{tabular}
		}
	\end{center}
\end{table*}
{}

\FigData{}

Molecular clouds have a complex structure, with filaments hosting dense
cores and immersed in a low density diffuse envelope. Large scale dust
continuum maps obtained with {\sl Herschel} have provided a
breakthrough, by showing the tight relationship between the filaments
and the dense cores.
These maps however do not provide information on the gas dynamics or its
chemical composition. Furthermore the relationship between the
submillimeter dust emission and the gas column density is affected by
the dust temperature and possible variations of the dust emissivity.
Molecular line emission maps provide alternative means to study
molecular cloud structure and relate it to the flow kinematics.
Molecular line emission is linked to the underlying physical properties
of the ISM, such as density, gas and dust temperatures, UV radiation
field, and cosmic ray ionization rate. But these relationships are
complex and their detailed study is a field in itself: astrochemical
modeling \citep{Le-Bourlot.2012,Agundez.2013}. Further complexity arises
when considering radiative transfer to derive line intensities from the
local chemical composition and physical structure.

The last few years have seen the installation of new wideband receivers
and spectrometer at millimeter and sub-millimeter radiotelescopes. With
these instruments, line surveys of several \GHz\ bandwidth and several
tens of thousands of spectral channels are the new default mode of
observations. Combined with wide field imaging capabilities both for
single dish and interferometers, hyperspectral imaging is now routinely
carried out with these instruments.

The analysis and interpretation of these large datasets, consisting of
thousands of spatial positions and tens of thousands of spectral
channels, will benefit from the use of statistical tools.
Principal Component Analysis (PCA) is one of the most widely used
multivariate analysis method. It has been used to study the Interstellar
Medium (ISM) using molecular emission maps
\citep{Ungerechts.1997,Neufeld.2007,Lo.2009,Melnick.2011,Jones.2012}.

In this paper, we address the following question: can PCA provide a
method to study the underlying physics of the ISM when applied to a
large dataset of molecular emission, without performing either radiative
transfer or astrochemical modeling? 

The article is divided as follows. Section~\ref{sec:data} presents the
data used in this study. Section~\ref{sec:methods} describes the
statistical method used in this paper and its implementation. Results
are presented in Sect~\ref{sec:results} first by analyzing the output of
the PCA, and further by comparing theses outputs with independent maps
of physical conditions in Orion\,B in Sect.~\ref{sec:correlations}. The
last section discusses these results.

\section{Data} 
\label{sec:data} 

The data used in this paper is selected from the ORION-B project (PI:
J.~Pety), which aims at mapping with the IRAM-30m telescope a large
fraction of the South-Western edge of the Orion\,B molecular cloud over
a field of view of 1.5 square degrees in the full 3\,mm atmospheric
window at 200 kHz spectral resolution. \citet{Pety.2016} describe in
detail the data acquisition and reduction strategies.
Table~\ref{tab:value} lists the 12 transitions selected in this paper
from the already observed frequency range (from 84 to 116\GHz), based on
the inspection of the full data cube.

For each line, we focus on emission coming from a limited
1.5\kms-velocity range centered on the peak velocity (\ie, 10.5\kms) of
the main velocity component along the line of sight. Averaging the three
0.5\kms\ velocity channels allows us to get a consistent dataset from
the radiative transfer and kinematics viewpoints. In particular, we
avoid the need to disentangle 1) the effects of hyperfine structures of
some lines, and 2) the complex velocity structure of the source
\citep{Orkisz.2016}. 

The observed field of view covers $0.81\deg\times1.10\deg$ and contains
the Horsehead nebula, and the \Hii\ regions NGC\,2023, NGC\,2024,
IC\,434, and IC\,435. The angular resolution ranges from 22.5 to
$30.5''$. The 12 resulting maps have a common pixel size of 9\arcsec{}
that corresponds to a Nyquist sampling for the highest frequency line
observed (\twCO\Jone\ at 115.27\GHz). The maps thus contains
$315\times420$ pixels. At a distance of $\sim 400\pc{}$
\citep{Menten.2007}, the maps give us access to physical scales between
$\sim$50\,mpc and 10\pc.

Figure~\ref{fig:data} shows the 12 maps of the resulting brightness
temperature multiplied by an ad-hoc factor in order that they can share
the same color look-up table, even though the intrinsic brightness
temperatures of the different lines differ by more than one order of
magnitude. The relative calibration of the different lines is excellent
because they were observed with the same telescope at almost the same
time, as the observed bandwidth was covered in only two frequency
tunings. The noise for each map, along with the minimum, median,
maximum, and variance values are listed in Table~\ref{tab:value}. The
noise is computed by fitting a gaussian function to the negative part of
the histogram of pixel brightnesses. This enables to compute the noise
without needing to mask out the emission.

\section{Principal Component Analysis} 
\label{sec:methods}
\subsection{Principle}

We use the following standard statistical terms: the dataset is composed
of \emph{samples} each described by individual \emph{features}.
In our case, each spatial pixel is a sample and each line intensity is a
feature (the full dataset thus corresponds to a data matrix of 132\,300
samples times 12 features).

Principal Component Analysis (PCA) is a widely used multi-dimensional
analysis technique \citep{Jolliffe.2002}, which can be defined in
several mathematically equivalent ways.
It aims at finding a new orthogonal basis of the feature space (whose
axes are called principal components, or PCs), so that for each $k$, the
projection onto the hyperplane defined by the first $k$ axes is optimal
in the sense that it preserves most of the variance of the dataset (or
equivalently that the error caused by this projection is minimal).
PCA thus defines successive approximations of the dataset by hyperplanes
of increasing dimension.

This is equivalent to the diagonalization of the covariance matrix, so
that the principal components are naturally uncorrelated. It can be
thought of as finding the principal axes of inertia of the cloud of
samples about their mean in feature space, and is thus a way to analyze
the covariance structure of the data. The principal components are
ordered by decreasing projected variance. As a result PC1 is the axis of
largest variance in the data. PC2 is then the axis of largest variance
\emph{at constant PC1} (orthogonal to PC1) and so forth. Neglecting the
axes of lowest variance then allows to define a low-dimensional
hyperplane in which the dataset is approximately embedded. An important
property to keep in mind is the linearity of PCA (it defines
low-dimensional hyperplanes, and not general low-dimensional
hypersurfaces).

A common variant\footnote{It actually goes back to one of the two
earliest descriptions of PCA: \citet{Hotelling.1933}} in the application
of PCA is to normalize the variations of the dataset around the mean by
the standard deviation of each features, before applying the PCA.
This amounts to diagonalizing the correlation matrix instead of the
covariance matrix.
The correlation-based variant allows to avoid having one feature
dominating the variance, and is indicated if the relative scales of the
features are not relevant for the purpose of the analysis.
As the relative intensity scales of the different molecules used here
are largely affected by properties (dipole moment, elemental
abundances,...) that are not relevant for our analysis of the chemical
variations across the map, we use here the correlation-based version of
PCA.

In this work, we used the PCA implementation available in the
\texttt{Python} package {\tt scikit-learn}~\citep{scikit-learn}, which
uses a singular value decomposition to compute the principal component
axes.

\subsection{Reparametrization of input data}

\FigAsinh{} 

\FigAsinhFunction{} 

\subsubsection{On the need of a reparametrization}

As seen in Table~\ref{tab:value}, some of the tracers have large
dynamical ranges (2 orders of magnitude for \thCO\Jone\ and \twCO\Jone).

Figure~\ref{fig:asinh_reparam} shows the histogram of the brightnesses
temperatures of two lines with contrasting behavior in our dataset,
namely \chem{^{13}CO(1-0)}, and \chem{N_2H^+(1-0)}. As the dynamic range
is large both in intensity and number of pixels per bin, these
histograms use the Bayesian Blocks algorithm (\citealt{Scargle.2013},
using here the \texttt{Python} implementation from AstroML, see
\citealt{AstroML,AstroMLtext}), which adapts the bin width to the
underlying distribution. While the histogram of the \chem{N_2H^+(1-0)}
is Gaussian to first order, the histogram of \chem{^{13}CO(1-0)}
exhibits heavy tails similar to power laws. As a result, extreme
intensity values might dominate the covariance structure of the data,
hiding the variations at the more common lower intensity values.

From the physical viewpoint, taking the logarithm of the brightness
temperature is also desirable. PCA is a linear technique which decompose
the data as a sum of uncorrelated components. Applying it to the
logarithm of the data allows a decomposition as a product of factors,
and thus describes the data structure in terms of ratios, products and
power laws, more adapted to the underlying radiative transfer and
chemical effects. Taking the logarithms of the data is the equivalent in
astrochemistry to doing color-color magnitude diagram analysis in
optical or UV studies. \citet{Pety.2016} show that the line integrated
brightness temperatures of our dataset is to first order correlated to
the column density of matter along the line of sight. We expect this
aspect to appear in our PCA analysis, and second order chemical
variations around this trend, which would be revealed by line ratios,
are thus better described as multiplicative --rather than additive--
factors.

\FigDataAsinh{}

\subsubsection{Impact of noise} The presence of noise however causes the
possibility of negative values in pixels where some lines are
undetected.
The logarithm transform cannot be applied to these negative noise
values.
In addition, as the logarithm stretches the lowest values compared to
the largest ones, it will also tend to stretch the positive noise values
of the undetected pixels, thus giving them more weight in the covariance
of the data.

There may be two different reasons for a non-detection: either the
measurement is not sensitive enough to detect the line or the region
just does not contain the species that emits the line. The latter case
happens in particular in \Hii\ regions (\eg, IC\,434), where
\twCO{}\Jone\ is photo-dissociated by the far UV photons. In this
particular case, we could remove from our dataset all the samples
(pixels) where no \twCO{}\Jone\ is detected. This just assumes that no
\twCO{}\Jone\ detections at high sensitivity imply the absence of
molecular gas.
However, this method is not generic. For instance, \chem{N_2H^+}\Jone\
is only detected in dense cores~\citep{Pety.2016}, and restricting
ourselves to these regions would drastically limit the scope of our
study.
If we wish to use this important tracer of the molecular gas while still
covering the range of different chemical regimes present in our full
map, we thus need to find an alternative. Moreover, in the 3\,mm band,
radio recombination lines, which emit in \Hii\ regions, could in
principle be added to our dataset to study the formation of molecular
gas.

Adding a thresholding step before taking the logarithm will only worsen
the scope of undefined values. While there are PCA methods \citep[see
e.g.][]{Ilin.2010} that can take missing datapoints into account, they
rely on the fact that these missing points have the same statistics than
the measured points (i.e., when a value is missing, it is independent of
the actual value). This is clearly not the case here as the missing
values (undetected lines) are missing because they are below the
sensitivity threshold. These are called censored values in statistics.
We thus search for a function that is linear around 0 (in the noise
dominated domain), and which is asymptotically equal to a logarithm for
large values (compared to the noise level). The inverse hyperbolic sinus
function, $\asinh(x)$, fulfills these conditions. We thus used the
following function to reparametrize the data before applying PCA
\begin{equation}
	T(x) = a \asinh(x/a), 
\end{equation}
where the
parameter $a$ is the typical value for which the function behavior
changes from a linear to a logarithmic regime (see
Fig.~\ref{fig:asinh_function}). 

The only free parameter in the method is the threshold $a$.
Appendix~\ref{app.asinh} discusses our choice, \ie{}, $a= 8\times0.08 =
0.64\K$ equal to 8 times the median noise of the dataset. In short, we
select the value of $a$ that maximizes the correlations of the first 3
principal components with independent known measurements of the column
density, volume density, and UV illumination (see
Section~\ref{sec:correlations}). This appendix also demonstrates that
our results are quite insensitive to the exact value of $a$. 

The right column of Fig.~\ref{fig:asinh_reparam} shows the result of the
asinh transformation on the intensity distributions of two transitions
representative of bright (\thCO\Jone) and weak (\chem{N_2H^+}\Jone)
averaged lines. In the case of the bright lines, the dynamic range is
drastically reduced with the heavy tail being transformed into a second
peak in the distribution but with no values above 3. In the case of
\chem{N_2H^+} the distributions before and after reparametrization are
very similar.
Figure~\ref{fig:data_asinh} shows the 12 maps of the molecular emission
after reparameterization by the asinh function, but before the
normalization step of the PCA analysis. The brightness temperatures of
all the maps have been compressed between about -0.5 and 3.5, with low
signal-to-noise brightness temperatures between -0.5 and 0.5 being
mostly untransformed.

\section{Results} 
\label{sec:results}

The PCA method exposes the correlations between the line brightness
temperatures. The derived PCs give the main axes of correlated
variations in the data set. As such, it does not directly yield physical
information underlying the dataset. In this section, we describe the
results of the application of this statistical method to our dataset and
we start to discuss their possible physic interpretation based on our
\emph{a priori} astrochemical knowledge. The possible relations between
these PCs and physical variables are investigated in a later section.

\subsection{Correlation fraction explained by the different principal
components} \FigExplainedVariance{} 

Figure~\ref{fig:explained_variance} shows the percentage of the
correlation explained by each PC (as a function of the principal
component number) along with the cumulative explained correlation as a
function of the number of principal components kept in the
decomposition. 

The first principal component explains the majority (60\%) of the total
correlation present in the original dataset.
Thus a large part of the variations in the dataset occur along a single
axis (i.e. all lines are strongly correlated to each other).
The second principal component accounts for about 10\% of the
correlation.
It is significantly less than the first component, but more than any
other components. PC 3, 4 and 5 correspond to similar amounts of
correlations (around 5\% each) and PC6 slightly less (3.3\%). PC1 to 6
collectively explain more than 90\% of the correlation in our dataset.
The remaining PCs have similar low amounts of explained correlation
(from 2\% for PC7 to 0.9\% for PC11).

\subsection{Discussion of the principal components}

\FigEigs{} 

\FigWheel{}

\FigProjMaps{}

The PCs defined by our analysis represent new axes in the feature space
(the full 12 PCs are simply a rotation of the initial basis of the
feature space), deduced from the data itself. They can thus be expressed
in terms of the original axes, as a linear combination of the
(transformed) line intensities. Figure~\ref{fig:eigs} displays the
quantitative contribution of each initial feature (line) to each PC.
{An alternative view of the relationship between the PCs and the
line intensities, namely the correlation wheels, is presented in Fig
~\ref{fig:wheel}.}

Each sample (pixel brightness) can then be projected on the new axes,
providing new coordinates commonly called \emph{component scores}. The
PCA method considers the pixels as independent samples, and thus ignores
the spatial structure of the molecular emission. It is nevertheless
possible to reconstruct the maps of the component scores.
Figure~\ref{fig:proj_maps} shows these projected maps. The chosen color
look-up table emphasizes that positive (red-colored) and negative
(blue-colored) values of the projected maps --corresponding to
variations above and below the average along the considered axis--
clearly extract a different spatial pattern per principal component.

The first principal component is a linear combination of all tracers,
with similar positive weights for all lines (Fig.~\ref{fig:eigs}). It
thus describes correlated variations of all molecules, and these account
for most of the variations in the dataset. It is a natural consequence
of the fact that all lines are well correlated (positively) to each
other.
\citet{Pety.2016} show that the emission of all lines is correlated to
first order with the column density of matter along the line of sight.
This component is thus probably related to the total column density,
whose increase causes to first order an increase in all lines (in the
linear approximation of PCA, non linear effects such as saturation of
the \twCO\ line are not captured). The corresponding component map
(Fig.~\ref{fig:proj_maps}) indeed resembles a map of column density.
This relation between PC1 and the total column density of matter will be
investigated more quantitatively in Sect.~\ref{sec:correlations}.
Note that as PC1 has only positive coefficients for all lines,
orthogonality ensures that all other PCs will represent contrasts
between different lines.

The second principal component represents the axis of largest variation
at constant PC1 (orthogonal to PC1). This axis of variations is
dominated by positive contributions of \chem{N_2H^+} and \chem{CH_3OH},
and negative contribution of \twCO\ and \thCO. The first two tracers are
chemically associated with dense and cold regions of the interstellar
medium. For instance, \chem{N_2H^+} is easily destroyed by CO, it can
thus only be abundant in the gas phase when CO has been depleted on the
grain surfaces~\citep{Pety.2016}. The component map shows strong
positive values highlighting known dense cores, including the clumps in
the head and in the neck of the Horsehead \citep{Ward-Thompson.2006} .

The third principal component shows positive contribution of CCH and CN
that are known to be sensitive to UV illumination, and negative
contribution of \chem{N_2H^+}, \chem{CH_3OH} and of the CO
isotopologues, that trace gas shielded from the UV field. This component
thus probably traces the chemical specificities of UV illuminated gas.
The component map clearly shows positive values at the eastern edge of
the cloud, illuminated by $\sigma$ Ori, and in the star-forming region
NGC\,2024.

The fourth principal component is particular in the fact that its map
almost only shows large (positive or negative) values in the regions of
large positive values of PC2. It thus highlight further chemical
variations inside dense cores. This component is completely dominated by
opposite contributions of \chem{CH_3OH} and \chem{N_2H^+} and thus
traces variations in the ratio of these two lines.
The component map seems to highlight smaller size cores embedded in some
of the clumps revealed by PC2, and thus probably highlights the
chemistry of the densest cores (larger \chem{N_2H^+} to \chem{CH_3OH}
ratios).

The fifth principal component shows positive contributions of sulfur
species (CS and SO), and \CeiO{}, and negative contributions of \twCO{},
CCH, and \chem{CH_3OH}. Its large positive values highlight larger scale
regions embedding the dense clouds shown by PC2 (but with negative
values where PC2 is very large), and this PC could thus trace the
chemistry of moderately dense gas.

The sixth principal component shows negative contributions of HCN,
HCO$^+$ and CN, which can all originate in photochemistry, and positive
contributions from CCH, \CeiO\, \thCO\ and SO, which are usually
associated with more shielded gas although CCH can also be bright in UV
illuminated regions. Its component map shows a wide blue region around
NGC\,2024, similar to the large warm dust region seen in the dust
temperature map of the region \citep{Schneider.2013}. It could thus also
be related to the radiation field, but trace a different aspect from
PC3, characterized by lower CCH intensities relative to the other lines.

The remaining components are more difficult to interpret, but tend to
describe opposite variations in pairs of lines that varied together in
previous PC. PC7, 9 and 10 display opposite variations in pairs of line
of the group HCN, HNC, CN and HCO$^+$, whose variations were correlated
in the previous PCs in which they had large weights (PC1 and 6). PC8
shows anticorrelated variations of SO and \CeiO, and its component map
shows a striking spatial pattern with negative values (high SO/\CeiO\
ratios) in the Horsehead, the molecular gas at the base of the
Horsehead, and the small scale clumps in NGC\,2024, and positive values
(low SO/\CeiO\ ratios) in a dense filament stretching away from
NGC\,2024.
PC11 is strongly dominated by CS, and thus shows specific variations of
CS, mostly uncorrelated with the other lines (somewhat anticorrelated
with \CeiO), and that were not described by the previous PCs. Its
component map shows small scale spots of positive values, mostly
surrounding NGC\,2024 and NGC\,2023. The fact that it appears so late in
the decomposition can be explained by the small size of the highlighted
region, thus having little weight in the correlation matrix. PC12 is
completely constrained by orthogonality to the previous PCs and is thus
only a consequence.

\subsection{Studying the effect of noise} 
\label{sec:noise}

The noisy nature of our data can have two kinds of effects. It can first
induce variability in our results (the results would vary for a
different realization of the random noise). {We verified the
stability of our results by using a bootstrapping method. Bootstrapping
is a method of choice to compute uncertainties on an estimator (here the
PCA components) when the distribution of estimator values cannot be
assumed to follow a simple distribution \citep{Feigelson.2012}. The idea
of bootstrapping is to use a Monte Carlo method to create new resampled
datasets of the same size as the original dataset by sampling with
replacement from the original dataset. We construct 5000 such
bootstrapped datasets and run the PCA algorithm on each. Because the PCA
is invariant through the change of sign of the PCs, we ensure that the
signs are consistent before computing the distribution of the PC
coefficients to avoid overestimating the uncertainty.
The results are presented both for the eigenspectra in
Fig.~\ref{fig:eigs}, and the correlation wheels in
Fig.~\ref{fig:wheel}.}

The PC coefficients appear overall very stable, the variances being
completely negligible for PC 1 and 2, and very small for PC 3 to 6. Only
PC 7 and 8 show significant variability, with some coefficients changing
sign. This can be understood as PCA results are particularly sensitive
to noise when two PCs correspond to very close eigenvalues, and PC 7 and
8 have the closest eigenvalues with respectively 2\% and 1.9\% of the
total correlation. Indeed, PCs with equal eigenvalues are degenerate in
the sense that any basis of the subspace they define satisfies the
definition of PCA. As a result, when eigenvalues are not exactly equal
but very close, the noise can result in a random rotation of this group
of PCs inside their subspace. Our results, which focus on the first few
PCs, are thus unaffected by noise variability.

The second possible effect of noise is to bias the results. PCA is
unbiased if the noise is spherical (i.e., has equal variance in all
directions) in the final dataset on which PCA is applied (i.e. after
standardization in our correlation-based variant). In this case, the
noise can only hide the lowest PCs (that describe variations smaller
than the noise level) and make them degenerate. In our case however, the
noise levels on the different molecular lines are initially close but
not equal (variations by a factor of 3 at most, see Table 1). The
non-linear reparametrization keeps these relative variances. Finally,
the last normalization step (by the standard deviation) gives final
noise variances proportional to the ratios of noise standard deviation
to total standard deviation of the reparametrized intensities. These
differ by up to a factor of 14.8 between the lines, and possible biases
may be present in our results, giving higher weight to the lines with
the largest ratios of noise variance to total variance. However, it
wasn't possible with the PCA method to both avoid giving higher weight
to the brightest lines (which led us to use the correlation-based PCA),
and at the same time ensure equal noise on all variables. We note that
the previous PCA studies of molecular clouds were less concerned by
noise-induced bias as they only used lines that were clearly detected in
all pixels. Our choice allowed us to perform the principal component
analysis on the full region, which led to the identification of two PCs
associated with dense core chemistry.

\section{Correlation of the principal component maps with
\emph{independently measured} physical parameters maps}
\label{sec:correlations}

\FigPhysicalParams{}

In the previous section, we combined two sources of information to
interpret the main principal components: 1) astrochemistry teaches us
that some molecules trace certain physical conditions, 2) Orion\,B is an
extremely well studied source, implying that the spatial structure of
the source is well known. For instance, the molecular cloud is known to
be illuminated by well-defined young massive stars \citep[see discussion
in][]{Pety.2016}. This allowed us to infer a link between the first
three principal components and physical quantities such as the column
density, the volume density, and UV illumination. In this section, we
will quantitatively assert these potential relations by studying the
correlation of each component map with a set of \emph{independently
measured} maps of physical parameters.

\subsection{Independent measure of the physical quantities}

The goal of this section is to find the principal component that is best
associated to each of the physical parameters, not to assign an absolute
physical meaning to some of the components. It is therefore not
necessary to have absolute values of the \emph{independently measured}
physical parameters maps. Only the relative variation of each physical
parameter is required to compute the correlation coefficient.
Figure~\ref{fig:physical_params} shows the different maps of the
physical quantities that we will correlate with the first 3 principal
components. This section describes how these maps were obtained.

\subsubsection{Column density} 

The dust column density map is from the Hershel Gould Belt Survey (PI:
P.Andre) Orion\,B map~\citep{Andre.2010,Schneider.2013}
\footnote{\url{http://www.herschel.fr/cea/gouldbelt/en/}}. This map was
obtained by fitting the far infrared Spectral Energy distribution by
greybodies. We apply a logarithmic scaling to the data to reduce the
dynamical range, the resulting maps is plotted in the left panel of
Fig.~\ref{fig:physical_params}.

\subsubsection{Volume density} 

Volume density is a difficult quantity to measure because one needs both
a mass estimate and an associated volume. Density is thus dependent on
the scale that it is computed at. We use the catalog of cores identified
and characterized in \citet{Kirk.2016}. We compute masses from each
cloud's 850\mum\ flux using their equation (3). To do this, we assume a
common temperature of 17\K\ for all clouds. From this mass and their
observed size estimates we compute a volume density for each of the
dense cores in our observed field of view. In this case correlation
cannot be carried out over the full map but we correlate the density
measured for each core with the value of the principal components
measured in the nearest pixel. The data is shown as a scatter plot in
the middle panel of Fig.~\ref{fig:physical_params}.

\subsubsection{UV radiation field} 

We compute the UV radiation field by using the fact that PAH emissivity
is roughly constant per unit H and unit radiation field
\citep{Draine.2007}. In practice, we use the WISE \citep{Meisner.2014}
12\mum\ maps divided by the column density clipped to a maximum value of
$\dix{22}\pscm$. We do not claim to have an absolute value of the UV
radiation field but a quantity that should be proportional to it. The
quantity $\log (U/\bar{U})$ where $\bar{U}$ is the mean value of $U$ is
shown in the rightmost panel of Fig.~\ref{fig:physical_params}.

The proper way to compute the UV radiation field from PAH emission would
be to divide by the \emph{volume density} but as we discussed in the
previous paragraph, it is not possible to get a \emph{full map} of
volume density. We choose to use column density as a proxy for volume
density even though it entails strong constrains on the spacial
distribution of the gas along the line of sight. Since we are interested
in relative variation of density and not absolute values it is
sufficient to assume that the matter is clustered into clouds that are
of similar spatial extents.

\subsection{Correlation of principal component maps with physical
parameters}

\begin{table}
	\caption{ 
	\label{tab:correl}
	Spearman rank correlation coefficient between the principal components
	and the physical quantities.} 
	\begin{center}
		\begin{tabular}
			{lrrrr}
			\hline  & {$\log N_{\mathrm{H_2}}$}  & {$\log n_{\mathrm{H}}$} & {$\log \left(U/\bar{U}\right)$} \\
\hline
PC1 & $ 0.90$ & $ 0.43$ & $-0.66$ \\ 
PC2 & $-0.57$ & $ 0.22$ & $ 0.43$ \\ 
PC3 & $-0.20$ & $ 0.06$ & $ 0.42$ \\ 
PC4 & $-0.01$ & $-0.23$ & $-0.04$ \\ 
PC5 & $-0.16$ & $ 0.02$ & $ 0.09$ \\ 
PC6 & $ 0.04$ & $-0.07$ & $-0.26$ \\ 
PC7 & $-0.02$ & $-0.03$ & $ 0.00$ \\ 
PC8 & $-0.02$ & $ 0.05$ & $ 0.12$ \\ 
PC9 & $-0.02$ & $ 0.06$ & $-0.11$ \\ 
PC10 & $-0.06$ & $-0.10$ & $-0.07$ \\ 
PC11 & $-0.03$ & $-0.13$ & $-0.04$ \\ 
PC12 & $ 0.04$ & $ 0.11$ & $-0.05$ \\ 
 \hline 
		\end{tabular}
	\end{center}
\end{table}
{} 

\FigPhysicalCorr{} 

We compute the Spearman rank correlation coefficient between each pair
of principal component maps and physical parameters maps. We use the
Spearman rank correlation instead of the Pearson linear correlation
coefficient because the potential relations between the principal
components and the physical parameters are most certainly non linear in
nature. The rank coefficient used is only sensitive to the ordering of
the values, and is thus not affected by the possible non-linearities of
the correlation.
Table~\ref{tab:correl} summarizes all theses values and
Fig.~\ref{fig:correl} shows the scatter plots for the most significant
correlations discussed in the next paragraphs. {An alternative
way of exploring the correlations between the independent physical
quantities and the Principal Components is to represent the correlation
between each physical parameters and the PCs in the correlation wheels
of Fig~\ref{fig:wheel}.}

For this analysis, it must be kept in mind that while the principal
components are necessarily uncorrelated, the physical parameters
considered here are correlated :
\NHt\ is an integral of \nH\ along the line of sight and the two are
thus strongly correlated, \U\ is inversely proportional to \NHt\ by
construction and they are thus anticorrelated.
As a result, the principal components will tend to represent the
uncorrelated part of the variations of the underlying physical
parameters.

\paragraph*{Column density:} The component maps showing the highest
correlation coefficient with \NHt\ is PC1. Spearman's rank correlation
coefficient is extremely high, \ie{} 0.90, and the scatter plot
(Fig.~\ref{fig:correl}, left panel) shows a strongly linear relation
between PC1 and $\log(\NHt)$.

As it is the first PC (axis of largest variation), it is unaffected by
the decorrelation constrain that affects the other PCs.
This first principal component can thus be interpreted as a global
measure of total column density, as suspected in our previous
discussion. Since \nH\ and \U\ are positively and negatively correlated
with \NHt, respectively, these physical parameters also exhibit
relatively strong positive, resp. negative correlations with PC1.

\paragraph*{Volume density:} The PC most correlated to \nH\ is also PC1,
due to the large correlation between \NHt\ and \nH.
The next principal components most correlated with our limited sample of
volume density measurements are PC2, which shows a Spearman's rank
correlation coefficient of 0.22, and PC4, with a Spearman's rank
correlation coefficient of -0.23.

As was discussed in Sect~\ref{sec:results}, PC2 and PC4 both trace
chemical differences typical of dense cores. PC2 and PC4 can thus be
interpreted as indicator of the presence of dense cores.
Note that this comparison was only done with a limited sample of rather
dense clouds.
We can thus only say that PC2 traces increased density among dense
clouds.
Because of the opposite sign of the correlation of the density with PC4,
negative values of this PC probably trace an even higher density regime.
As noted before, the behavior of PC2 in less dense region is probably
anti-correlated with density, and these PCs are thus only indicative of
density in the high density regime.

\paragraph*{Radiation field:} For the radiation field, the most
correlated PC is again PC1 (negative correlation) as high column density
tends to result in highly shielded gas. Not considering PC1 and PC2, the
third principal component shows the highest correlation with our
estimation of the radiation field, with a Spearman's rank coefficient of
0.42. It thus describes the part of the radiation field variations that
are not correlated with the cloud column density. As a result, it
highlights the part of the cloud where specific sources cause increased
illumination. A strong positive correlation (0.43) with PC2 is also
found, which is most likely an artifact due to the positive values of
PC2 in the diffuse regions surrounding the molecular cloud (where most
of the lines involved in PC2 are undetected, making PC2 irrelevant). PC6
also has a significant correlation with the radiation field (-0.26).
These results thus confirm our previous discussion of PC3 and PC6.

{These results can be inferred graphically from the correlation
wheels of Fig~\ref{fig:wheel}, where the colored arrows tracing the
location of \NHt\ (red), \nH\ (green) and \U\ (blue) the PC space have a
significant size only for the first 4 PCs.
Furthermore, each arrow is roughly aligned with one of the PC : PC1 for
\NHt, PC2 for \nH\ and PC3 for \U.}

\section{Discussion} 
\label{sec:discussion} 
\subsection{Comparison with
other works}

PCA has been extensively used in astronomy as a multivariale analysis
tool starting from the work of \citet{Deeming.1964} on the
classification of stellar spectra. Its use for the study of molecular
maps of the ISM is more recent starting with the work of
\citet{Ungerechts.1987}. Notable studies include
\citet{Neufeld.2007,Lo.2009,Melnick.2011}, and \citet{Jones.2012}

We first discuss the common points between these studies. On a technical
aspect, all these studies apply only subtraction by mean and
normalization by variance and do not attempt to introduce a non linear
reparametrization of the observed intensities. The effect of noise is
considered by limiting the number of observed lines to the set of
brightest tracers \citep{Lo.2009} and masking regions of low emission
\citep{Jones.2012}. All of these studies identify the utility of PCA as
a means of studying the correlations between molecular lines by studying
the commonality between tracers and its variation, and as a tool to
identify regions interesting for further study. With the notable
exception of \citet{Ungerechts.1987} rarely a discussion is made
relating the principal components with the underlying physical
parameters of the ISM although often, specific correlations or
anticorrelations are discussed in a chemical view or by invoking opacity
effects \citep{Lo.2009}.

\citet{Ungerechts.1997} present a dataset of 360 spatial points in 32
lines of 20 chemical species including isotopologues toward the Orion\,A
molecular cloud with the 14m FCRAO telescope. Using a Principal
Component Analysis (PCA) they show that the chemical abundances of most
species stay similar for the Orion ridge, and that the main differences
stand up for the BN-KL region. They note that their first 3 PCs contain
80\% of the observed correlation and use the component maps mainly to
identify regions for further astrochemical study. They nevertheless
discuss that data mostly lie in a 3D space spanned by the first three PC
because the molecular emission probably depends on three physical
parameters of the ISM the column density, volumic density and gas
temperature. \citet{Melnick.2011} compared the distribution of the
ground state transition of water vapor with that of the ground state
transition of N$_2$H$^+$, CCH, HCN, CN, and $^{13}$CO(5-4). Water vapor
is found to best correlate with species like $^{13}$CO(5-4) and CN,
tracing the cloud surface up to a few magnitudes of extinction and is
poorly correlated with N$_2$H$^+$ tracing the shielded regions. Using
MOPRA, \citet{Jones.2012} have mapped the central molecular zone (CMZ)
near the center of the Galaxy in 20 spectral lines in the 85.3 to
93.3~GHz range. They performed a PCA analysis using the strongest eight
lines (HCN, HCO$^+$, HNC, HNCO, N$_2$H$^+$, SiO, CH$_3$CN and HC$_3$N)
in the restricted area around SgrB2 and SgrA where the N$_2$H$^+$ line
is stronger than 10\Kkms. The analysis recovers the overall similarity
of the line maps. The main differences are found in the SgrA and SgrB2
cores between the bright lines HCN, HNC and HCO$^+$ and the other
species, and is attributed by \citet{Jones.2012} to a difference in
opacity. The other PCA components reveal specific regions where
CH$_3$CN, HNCO and SiO abundances are enhanced, possibly due to shocks
or hot cores. \citet{Lo.2009} studied the G333 molecular cloud with
MOPRA. The PCA is performed on eight molecular lines with high S/N ratio
($^{13}$CO, C$^{18}$O, CS, HCO$^+$, HCN, HNC, N$_2$H$^+$, CCH). The PCA
analysis reveals differences between the regions traced by CCH and
N$_2$H$^+$, and in star forming regions an anticorrelation between
$^{13}$CO, C$^{18}$O and N$_2$H$^+$, and between N$_2$H$^+$ and HCO$^+$.
PCA was also used by \citet{Neufeld.2007} to separate the different
regions impacted by supernovae shock waves.

While the previous analysis used integrated line intensities, PCA has
also been used on spectral line profiles as a mean to extract
information on the spatial properties of the turbulence
\citep{Heyer.1997,Roman-Duval.2011,Brunt.2013}, to study line absorption
depth \citep{Neufeld.2015}, or to measure cloud properties
\citep{Rosolowsky.2006}. To our knowledge no PCA analysis takes into
account the full velocity profile of the molecular emission at every
spatial position. Further inquiry on this subject is required as it can
add a further dimension (the shape of the line profiles) to study the
emission correlations.

\subsection{Non-linearities and multiple physical regimes}

Two important properties of PCA must be kept in mind. The first is that
it is a linear method. It distinguishes the axes of variations in the
dataset as linear combinations of the initial variables. Thus,
non-linear (approximate) relations between the variables cannot be
properly captured.
In this case, a single relation could be described by several PCs, one
describing the best linear approximation, and additional PCs describing
directions in which the non-linear relation deviates from linearity.
Using our non-linear transform, equivalent to a logarithm for high SNR
values, alleviates part of the problem as it allows to describe
power-law relations, but other non-linearities (such as saturation for
the \twCO\ line) are still not captured. 

Non-linear extensions of PCA exist (kernel-PCA, neural network-based
dimensionality reduction such as the Self-Organizing Maps), but their
results tend to be harder to interpret.

The second property is that PCA is based on the global correlation
matrix of the data. If different physical regimes are present in the
dataset, each with different relations between the variables, PCA will
do a global (linear) approximation over the different regimes. Depending
on the fraction of samples (pixels) representing the different regimes,
it may give more weight to some regimes than others, neglect some
regimes, or mainly represent one of the regimes.

\subsection{Reduction of dimensionality}

PCA is often used as a dimensionality-reduction tool (approximating the
data by its projection on a lower dimensional hyperplane in feature
space), by keeping only a subset of the PCs that account for a
sufficiently large fraction of the variance in the dataset.
We saw that PC1 to 6 explain more than 90\% of the correlation structure
of the data.
Moreover, PC6 define a transition between several PC with similar levels
(5\% for PC3-4-5, 1-2\% for the PCs after 6).
The projection on the first 6 PC thus define a 6-dimensional hyperplane
in which the data is approximately embedded.
However, we saw some striking spatial features appearing in later PCs,
indicating meaningful axes of variations, such as PC8 and 11. The
pattern of small scale spots shown by PC11 (corresponding to overbright
CS) is particularly interesting, and its late apparition in the
decomposition could simply be a consequence of the small fraction of
pixels concerned.
Thus, even PCs with lower fraction of explained correlation can contain
important information, such as specific variations occurring in small
regions only.

\subsection{A synthetic view of Orion\,B} 

\FigColor{} 

Using the physical interpretation of the principal components derived
from the previous section it is possible to derive a synthetic view of
the Orion\,B cloud rendered through a color image (see
Fig.~\ref{fig:color}). The principal components 1 (column density), 2
(density) and 3 (UV illumination) are used in the following way: the
column density is used to encode the luminosity, and the density and UV
illumination are combined orthogonally to define a color.
\begin{equation}
	\emr{hue} = \emr{atan}(uv,density) 
\end{equation}
In
this way, it is possible to identify by color only, the physical
properties associated with every line of sight.

Most of the region is composed of low density gas either obscured
(green) or UV illuminated (yellow). Notable features are the moderately
dense (orange) photodissociation regions that are present as the surface
of pillars (e.g. around the Horsehead nebula) and as globules
surrounding the NGC\,2024 massive \Hii\ region in the upper part of the
map. A sharp illumination gradient is visible at the base of the neck of
the Horse with transition from illuminated (yellow) to shielded (green)
gas.

Concerning the dust lane in front of NGC\,2024, there is a clear sharp
frontier between the northern and southern part, the north being
strongly UV illuminated (yellow, orange, and red), the southern part
much more obscured (cyan and green). The variation in the density of
dense cores are visible with transitions from moderately dense (cyan) to
higher density (dark blue) gas.

\section{Conclusion}

In order to study the correlations between maps of the emission of 12
bright lines belonging to the 3\,mm band over the south-western edge of
the Orion\,B molecular cloud, we applied the principal component
analysis to these data. Before this analysis, we applied a non-linear
transformation that is close to linear around zero and is equivalent to
a logarithmic transform at large values. The goal of this non-linear
transform is two-fold. 1) Ratios of brightness temperatures are easier
to interpret but PCA assumes that the relations in the input data set
are linear.
Applying the logarithm to the input data allows us to transform ratios
of brightness temperatures into subtractions well adapted to a linear
analysis. 2) Signal is only detected on a line-dependent subset of the
field of view.
Applying the logarithm to noisy brightness temperatures centered around
zero is mathematically ill-defined. Having a linear transform around
zero solves this problem. We tuned the transition value between the
linear and logarithmic value that is typically 8 times the typical noise
value of the dataset. We showed that the results are not very sensitive
to this value.

The PCA delivers a set of maps that are linear combination of the input
brightness temperatures, taking into account their
\mbox{(anti-)correlations}.
While PCA does not use the spatial information of the input dataset, the
output maps expose well-defined structures.
We thus limited our analysis to the first few principal components that
expose the largest correlations present in the initial dataset. The
analysis of these correlations allowed us to propose some link between
the first three components and physical parameters, as the column
density, volume density, and UV radiation field. We quantified these
links by computing the correlation coefficients of these principal
components with independent measurements of the column density, volume
density, and UV illumination. The first principal component is highly
correlated to the column density measured from the dust extinction and
has positive contributions from all molecules, as has been noted in
\citet{Pety.2016}. The third principal component is well correlated to
our estimation of the UV illumination, with positive contributions from
CCH, CN and anticorrelations with \chem{N_2H^+} and \chem{CH_3OH}. The
second principal component is correlated with the volume density in the
dense cores with a combined positive contribution with \chem{N_2H^+} and
\chem{CH_3OH} and a negative contribution from \twCO\ and \thCO.

The possibility to link linear combinations of the brightness
temperatures of a set of 3\,mm lines to physical quantities as the
column density, volume density, or UV illumination opens an interesting
avenue to analyze the large spectro-imaging data sets that (sub)-mm
radioastronomy starts to produce.
As PCA analysis only works on the brightness temperatures independent of
their spatial relations, it also offers an easy possibility to compare
with large grids of detailed 1D models of photo-dissociation regions. In
future papers, we will continue to explore this with more advanced
decomposition techniques that may take into account missing values,
noise effects, or non-linear relations in the input dataset.

\acknowledgements{This work was supported by the French program Physique
et Chimie du Milieu Interstellaire (PCMI) funded by the Conseil National
de la Recherche Scientifique (CNRS) and Centre National d'Etudes
Spatiales (CNES). We thank the CIAS for its hospitality during the two
workshops devoted to this project. PG thanks ERC starting grant (3DICE,
grant agreement 336474) for funding during this work. PG's current
postdoctoral position is funded by the INSU/CNRS. NRAO is operated by
Associated Universities Inc. under contract with the National Science
Foundation. We thank P.Andre and N.Schneider to kindly give us access to
the Herschel Gould Belt Survey data. This research has made use of data
from the Herschel Gould Belt survey (HGBS) project
(http://gouldbelt-herschel.cea.fr). The HGBS is a Herschel Key Programme
jointly carried out by SPIRE Specialist Astronomy Group 3 (SAG 3),
scientists of several institutes in the PACS Consortium (CEA Saclay,
INAF-IFSI Rome and INAF-Arcetri, KU Leuven, MPIA Heidelberg), and
scientists of the Herschel Science Center (HSC). We than the anonymous
referee for his/her constructive comments.}

\bibliographystyle{aa}

%\bibliography{pca-in-OrionB}
\bibliography{/Users/gratier/Documents/Biblio/biblio}

\appendix{}

\section{Optimal value of $a$ in the $\asinh$ reparametrisation}
\label{app.asinh}

\FigOptimalK{} 

The only free parameter in the $\asinh$ reparametrization is $a$, the
parameter which marks the boundary between the linear and logarithmic
regimes of the $\asinh$ function (see Fig.~\ref{fig:asinh_function}). As
shown in Table~\ref{tab:value} the noise across different lines is
similar and we express $a$ as the product of a constant factor $K$ by
the median noise $0.08\unit{K}$. The quantity we choose to maximize is
the mean of the absolute value of the correlation coefficient of the
principal components with the physical maps \NHt, \U, and \nH, we note
this quantity $\langle|\rho|\rangle$.
Figure~\ref{fig:optimalK} shows the evolution of this quantity with
increasing values of $K$, a maximum value of $\langle|\rho|\rangle$
around $K=8$ although an acceptable range of $K$ values (reduction of
$\langle|\rho|\rangle$) by less than 5\%) spans values from 3 to 40.

\end{document}